\documentclass{osa-article}

\newcommand{\be}{\begin{equation}}
\newcommand{\ee}{\end{equation}}
\newcommand{\pp}{\partial}
\newcommand{\vv}[1]{\boldsymbol{\mathrm{#1}}}

\usepackage{color}

\journal{boe}

\articletype{Research Article}

\begin{document}

\title{Fast and robust reconstruction algorithm for fluorescence diffuse optical tomography assuming a cuboid target}

\author{Chunlong Sun,\authormark{1,2} Yu Jiang,\authormark{3} Jijun Liu,\authormark{1} Manabu Machida,\authormark{4,*} Gen Nakamura,\authormark{2} and Goro Nishimura\authormark{5}}

\address{\authormark{1}School of Mathematics, Southeast University, 
Nanjing 210096, P. R. China\\
\authormark{2}Department of Mathematics, Hokkaido University, 
Sapporo 060-0810, Japan\\
\authormark{3}School of Mathematics, Shanghai University of Finance and Economics,
Shanghai 200433, P. R. China\\
\authormark{4}Institute for Medical Photonics Research, Hamamatsu University School of Medicine,
Hamamatsu 431-3192, Japan\\
\authormark{5}Research Institute for Electronic Science, Hokkaido University,
Sapporo 060-0810, Japan
}

\email{\authormark{*}machida@hama-med.ac.jp} 



\begin{abstract}
A fast and robust algorithm for fluorescence diffuse optical tomography is proposed. We identify the location of a fluorescence target by assuming a cuboid. The proposed numerical method is verified by a numerical experiment and an in vivo experiment.
\end{abstract}

\section{Introduction}

Fluorescence diffuse optical tomography (FDOT) is one type of optical tomography which makes use of fluorescence light from fluorophore. In FDOT, diffuse light from fluorophore such as Indocyanine Green (ICG) is detected on the boundary of biological tissue to obtain tomographic images \cite{Ntziachristos-Bremer-Weissleder03,Ntziachristos-Ripoll-Wang-Weissleder05}. FDOT has been verified in vivo \cite{Koenig-etal08,Ntziachristos-etal02} and also in clinical research for breast cancer \cite{Corlu-etal07}.

To obtain reconstructed images with good quality, a large number of source-detector pairs is necessary. The number of source-detector pairs can be increased by time-dependent experiments \cite{Gao-Zhao-Zhang-Tanikawa-Marjono-Yamada08}. The superiority of the time-resolved approach over the continuous-wave approach in FDOT was concluded \cite{Ducros-DAndrea-Bassi-Peyrin11} although the quality of images can be improved even for a time-independent experiment if a large data set is used \cite{Panasyuk-Wang-Schotland-Markel08}.

Even if many sources and detectors are used, the resolution of tomographic images from optical tomography cannot be compared to that from X-ray CT. In this paper, we assume a cuboid or rectangular parallelepiped in the medium. Then we try to find the cuboid instead of giving up reconstructing the shape of the target. Thus we can develop a fast and robust numerical algorithm using a cuboid.

The paper is organized as follows. In Sec.~\ref{eqs}, we develop the formulation of our FDOT and give an analytical formula for the emission light. The proposed numerical scheme is described in Sec.~\ref{cuboid} with a numerical example of an ellipsoidal target. In Sec.~\ref{beefexp}, we validate our numerical method with a beef experiment. Finally, Sec.~\ref{concl} is devoted to conclusion.

\section{Formulation}
\label{eqs}

Let us suppose that a fluorescence target is embedded in biological tissue occupying the half space ($-\infty<x<\infty$, $-\infty<y<\infty$, $0<z<\infty$). Let $c$ be the speed of light in the medium. Let $u_e(\vv{r},t)$, $u_m(\vv{r},t)$ be the energy densities of the excitation light and emission light, respectively. Here, $\vv{r}=(\vv{\rho},z)$ with $\vv{\rho}=(x,y)$ is position and $t$ is time. We assume that the reduced scattering coefficient $\mu_s'$ and the absorption coefficient $\mu_a$ are constants everywhere in the medium. In the medium ($z>0$), $u_e(\vv{r},t)$ and $u_m(\vv{r},t)$ obey the following diffusion equations.
\be
\left(\frac{1}{c}\frac{\pp}{\pp t}-D\Delta+\mu_a\right)u_e=0,\quad
\left(\frac{1}{c}\frac{\pp}{\pp t}-D\Delta+\mu_a\right)u_m=F,
\ee
where $D=1/(3\mu_s')$ and
\be
F(\vv{r},t)=
\frac{n(\vv{r})}{\tau}\int_0^te^{-(t-s)/\tau}u_e(\vv{r},s)\,ds.
\ee
Here, $n(\vv{r})$ is proportional to the fluorophore concentration and $\tau$ is the fluorescence lifetime. We assumed that $u_e=0$ and $u_m=0$ at $t=0$ but the sample is illuminated at position $\vv{r}_s=(\vv{\rho}_s,0)$ by a pencil beam of the temporal profile $f(t)$ ($t>0$) in the $x$-$y$ plane. Thus, $u_e,u_m$ satisfy the Robin boundary conditions at $z=0$ as
\be
-\frac{\pp}{\pp z}u_e+\beta u_e=f(t)\delta(\vv{\rho}-\vv{\rho}_s),\quad
-\frac{\pp}{\pp z}u_m+\beta u_m=0.
\ee
The parameter $\beta$ is given by $\beta=\frac{1}{2D}(1-2\int_0^1R(\mu)\mu\,d\mu)/(1+\int_0^1R(\mu)\mu^2\,d\mu)$ with the Fresnel reflectance $R(\mu)$, which  depends on the refractive index of the medium. Suppose the out-going light is detected at $\vv{r}_d$ on the boundary. The excitation light and emission light are detected through a response function $R$ as
\be
U_e(\vv{r}_d,t;\vv{r}_s)=\int_0^tR(t-s)u_e(\vv{r}_d,s)\,ds,\quad
U_m(\vv{r}_d,t;\vv{r}_s)=\int_0^tR(t-s)u_m(\vv{r}_d,s)\,ds,
\ee
where the function $R$ is determined by the detector. These $U_e,U_m$ correspond to experimentally measured light $U_e^{\rm exp}(\vv{r}_d,t;\vv{r}_s)$, $U_m^{\rm exp}(\vv{r}_d,t;\vv{r}_s)$.

Let $G(\vv{r},\vv{r}';t)$ be the Green's function which satisfies $(\frac{1}{c}\frac{\pp}{\pp t}-D\Delta+\mu_a)G(\vv{r},\vv{r}';t)=\delta(\vv{r}-\vv{r}')\delta(t)$, with the Robin boundary condition $-\frac{\pp}{\pp z}G+\beta G=0$. The Green's function is obtained as $G(\vv{r},\vv{r}';t)=c(4\pi Dct)^{-3/2}e^{-\mu_act}\exp(-\frac{(x-x')^2+(y-y')^2}{4Dct})g(z,z';t)$, where
$g(z,z';t)=\exp(-\frac{(z+z')^2}{4Dct})+\exp(-\frac{(z-z')^2}{4Dct})-2\beta\sqrt{\pi Dct}\exp\left(\beta(z+z')+\beta^2Dct\right)\mathop{\mathrm{erfc}}(\frac{z+z'+2\beta Dct}{\sqrt{4Dct}})$. We introduce
\be
q(t)=\int_0^tR(t-s)f(s)\,ds.
\ee
This $q(t)$ is called the instrument response function.

Let $\Omega_c$ be a cuboid specified by $x_1,x_2,y_1,y_2,z_1,z_2$ as $x_1<x<x_2$, $y_1<y<y_2$, $z_1<z<z_2$. We assume that
\be
n(\vv{r})=\left\{\begin{aligned}
M,&\quad\vv{r}\in\Omega_c,
\\
0,&\quad\vv{r}\notin\Omega_c,
\end{aligned}\right.
\ee
where $M>0$ is a constant. Then we obtain
\be
U_m(\vv{r}_d,t;\vv{r}_s)=
\int_0^tQ(s)\int_0^{t-s}\int_{\Omega_0}n(\vv{r}')
G(\vv{r}_d,\vv{r}';t-s-s')G(\vv{r}',\vv{r}_s;s')\,d\vv{r}ds'ds.
\label{Umtheor}
\ee
where we introduced $Q(t)=\frac{D}{\tau}\int_0^te^{-t'/\tau}q(t-t')\,dt'$. Equation (\ref{Umtheor}) can be rewritten as
\be
U_m(\vv{r}_d,t;\vv{r}_s)=
M\int_0^tQ(s)\int_0^{t-s}f_1(\vv{\rho}_d,\vv{\rho}_s,t-s,t';x_1,x_2,y_1,y_2)
f_2(t-s,t';z_1,z_2)\,dt'ds,
\label{Umtheor2}
\ee
where
\be
\begin{aligned}
&
f_1(\vv{\rho}_d,\vv{\rho}_s,t,s;x_1,x_2,y_1,y_2)
=
\frac{e^{-\mu_act}}{4^3\pi^2D^2t\sqrt{(t-s)s}}
e^{-\frac{(x_d-x_s)^2+(y_d-y_s)^2}{4Dct}}
\\
&\times
\left[h(x_d,x_s,t,s;x_2)-h(x_d,x_s,t,s;x_1)\right]
\left[h(y_d,y_s,t,x;y_2)-h(y_d,y_s,t,s;y_1)\right],
\end{aligned}
\ee
with $h(x_d,x_s,t,s;x)=\mathop{\mathrm{erf}}\left(\sqrt{\frac{t}{4Dc(t-s)s}}\left(x-\frac{sx_d+(t-s)x_s}{t}\right)\right)$
and $f_2(t,s;z_1,z_2)=\int_{z_1}^{z_2}g(0,z';t-s)g(z',0;s)\,dz'$. We can compute $U_m(\vv{r}_d,t;\vv{r}_s)$ using (\ref{Umtheor2}).

\section{Identification of a cuboid}
\label{cuboid}

To illustrate our proposed inversion scheme, we will reconstruct $n(\vv{r})$ using solely $U_m^{\rm exp}$. The reconstruction will be done with the Levenberg-Marquardt method \cite{Levenberg44,Marquardt63}. The choice of initial guesses is important.

Let $N_{\rm SD}$ be the number of source-detector pairs. The positions of each source and detector are denoted by $\vv{r}_s^i,\vv{r}_d^i$ ($i=1,\dots,N_{\rm SD}$). At each detector, light is measured at $N_t$ temporal points.

We first look for a rough location of the target in the $x$-$y$ plane or a region of interest $\Gamma$ on the boundary, under which the target is likely to be embedded, by observing $I_i=\int_0^TU_m^{\rm exp}(\vv{r}_d^i,t;\vv{r}_s^i)\,dt$ ($i=1,\dots,N_{\rm SD}$). We call this step of setting $\Gamma$ the topography process. Then we assume a cubic target whose location and size are determined by four parameters $x_0,y_0,z_0,l$ such that $x_1=x_0-l/2$, $x_2=x_0+l/2$, $y_1=y_0-l/2$, $y_2=y_0+l/2$, $z_1=z_0-l/2$, $z_2=z_0+l/2$. Choosing the initial guess for $x_0,y_0$ inside $\Gamma$, we solve the inverse problem of determining $x_0,y_0,z_0,l,M$ by the Levenberg-Marquardt method and obtain $\vv{a}_*^{\rm cubic}=(x_0',y_0',z_0',l',M')$. We refer to this step as the cubic tomography. Finally with the obtained values $\vv{a}_*^{\rm cubic}$ as the initial guess $\vv{a}_0^{\rm cuboid}$, we solve the inverse problem of determining a cuboid target. In this way, we can identify $\Omega_c$ for $n(\vv{r})$ by obtaining reconstructed values $\vv{a}_*^{\rm cuboid}=(x_1^*,x_2^*,y_1^*,y_2^*,z_1^*,z_2^*,M^*)$. This last step is called the cuboid tomography. The algorithm is summarized as follows.

\begin{description}
\item[Step 1.] (Topography process) Find $\Gamma$ on the boundary.
\item[Step 2.] (Cubic tomography) By searching underneath $\Gamma$, obtain reconstructed values $\vv{a}_*^{\rm cubic}$.
\item[Step 3.] (Cuboid tomography) Find $\vv{a}_*^{\rm cuboid}$ starting with $\vv{a}_*^{\rm cubic}$.
\end{description}

Below, we will demonstrate our algorithm with a numerical experiment. We set $\tau=0$, $f(t)=R(t)=\delta(t)$. Let us assume an ellipsoidal target $\{x^2/1.5^2+y^2/3^2+(z-11)^2/1.5^2\le1\}$, where the unit of length is ${\rm mm}$, with $n(\vv{r})=0.02\,{\rm mm}^{-1}$. In the numerical experiment, $5\%$-noise is added and $U_m^{\rm exp}=U_m(1+0.05\varepsilon)$, where $\varepsilon$ is drawn from the standard Gaussian distribution. We use the following $N_{\rm SD}=32$ source-detector pairs $p_i=(\vv{r}_s^i,\vv{r}_d^i)$ ($i=1,\dots,N_{\rm SD}$) on the boundary: 
$p_{4j-3}=(\xi_j,\zeta_j+10\sqrt{3};\xi_j-10,\zeta_j)$, $p_{4j-2}=(\xi_j,\zeta_j+10\sqrt{3};\xi_j+10,\zeta_j)$, $p_{4j-1}=(\xi_j,\zeta_j-10\sqrt{3};\xi_j-10,\zeta_j)$, $p_{4j}=(\xi_j,\zeta_j-10\sqrt{3};\xi_j+10,\zeta_j)$, $j=1,\dots,8$, where $(\xi_1,\zeta_1)=(-10,10)$, $(\xi_2,\zeta_2)=(-10,0)$, $(\xi_3,\zeta_3)=(-10,-10)$, $(\xi_4,\zeta_4)=(0,-10)$, $(\xi_5,\zeta_5)=(10,-10)$, $(\xi_6,\zeta_6)=(10,0)$, $(\xi_7,\zeta_7)=(10,10)$, $(\xi_8,\zeta_8)=(0,10)$. We set $N_t=20$ and $t_k^i=t_0^i+(k-1)\Delta t$ ($k=1,\dots,N_t$), where $\Delta t=6.67\,{\rm ps}$. Here, $t_0^i=(k_0^i-10)\Delta t$, where $U_m^{\rm exp}(\vv{r}_d^i,t;\vv{r}_s^i)$ has the peak at $(k_0^i-1)\Delta t$ ($i=1,\dots,N_{\rm SD}$). Thus we have $N_{\rm SD}N_t$ ($=640$) measured values $U_m^{\rm exp}(\vv{r}_d^i,t_k^i;\vv{r}_s^i)$ ($i=1,\dots,N_{\rm SD}$, $k=1,\dots,N_t$). Let $\vv{a}$ be a vector which contains unknown parameters to be reconstructed. We define
\be
F(\vv{a})=\sqrt{
\sum_{i=1}^{N_{\rm SD}}\sum_{k=1}^{N_t}\left(
U_m(\vv{r}_d^i,t_k^i;\vv{r}_s^i)-U_m^{\rm exp}(\vv{r}_d^i,t_k^i;\vv{r}_s^i)
\right)^2}.
\ee
By using the Levenberg-Marquardt method, we try to find $\vv{a}=\vv{a}_*$ that minimizes $F(\vv{a})$.

(Step 1) We find $I_{4}=I_{10}=I_{17}=I_{27}=5.9\times10^{-7}$, whereas $I_{1}=I_{11}=I_{20}=I_{26}=1.0\times10^{-9}$. Hence we can set $\Gamma=\{-10<x_0<10,-10<y_0<10\}$. At this moment, we have no knowledge about the depth at which the target is embedded. We suppose $0<z_0<30$ together with $0<l<\min(20,2z_0)$ and $0<M<10$.

(Step 2) We set the initial values of $x_0,y_0$ in $\Gamma$. For example, we can start from $\vv{a}_0^{\rm cubic}=(x_0,y_0,z_0,l,M)=(2,2,5,4,0.1)$. The reconstructed values are
\be
\vv{a}_*^{\rm cubic}=(x_0',y_0',z_0',l',M')=
(0.0,\,0.0,\,11.24,\,4.089,\,0.0086).
\label{cubicrecon}
\ee
We note that choosing $x_0,y_0$ in $\Gamma$ is important for fast convergence. If we set $(x_0,y_0)=(-15,-15)$, more than $100$ iterations are needed whereas about $10$ iterations are sufficient for suitable $x_0,y_0$.

\begin{figure}[htbp]
\centering
\includegraphics[width=0.4\textwidth]{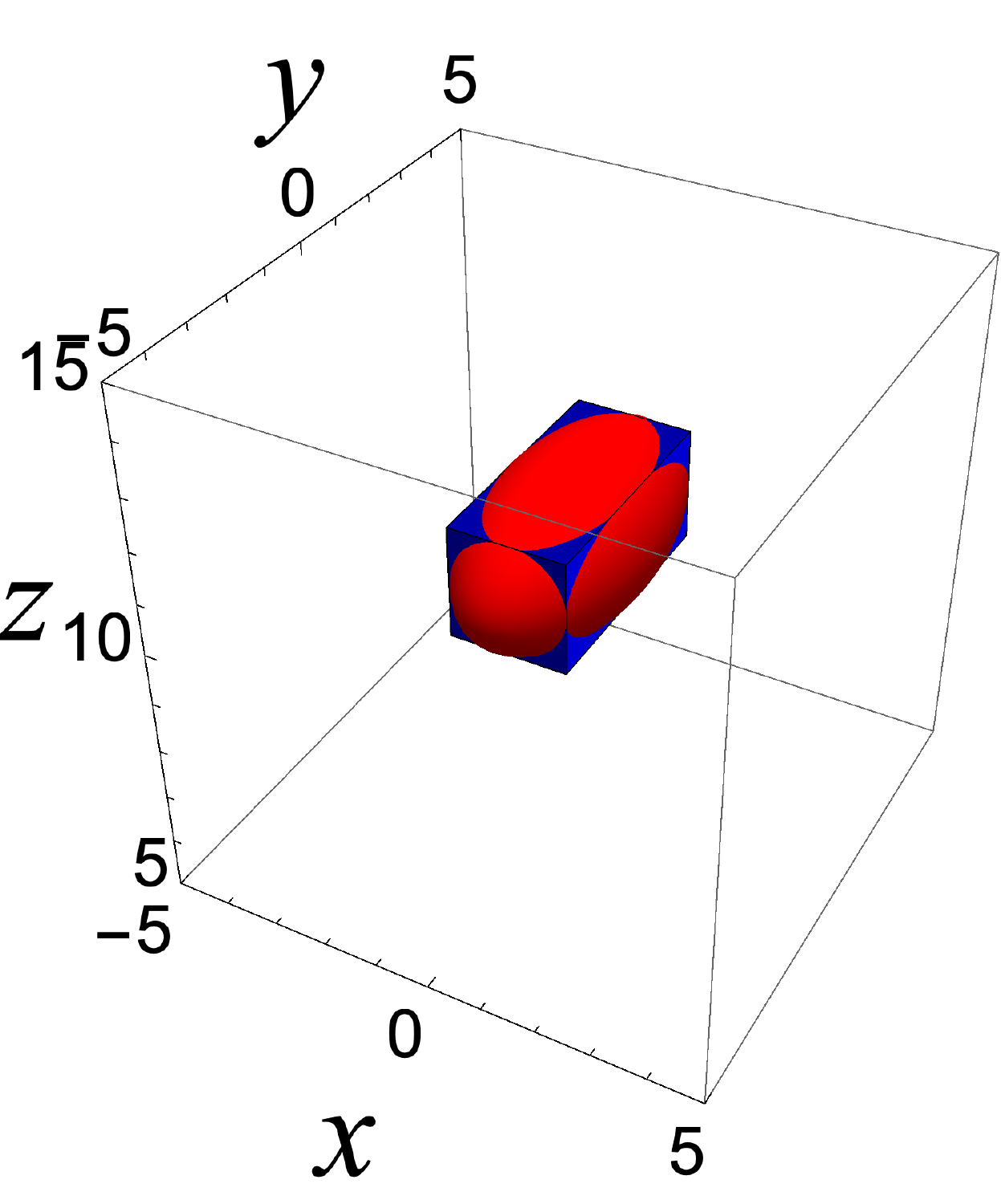}
\caption{
The identification of the ellipsoidal target by a cuboid. The obtained cuboid given by $\vv{a}_*^{\rm cuboid}$ (blue) is shown with the ellipsoid (red).
}
\label{fig:ellipsoid1}
\end{figure}

\begin{figure}[htbp]
\centering
\includegraphics[width=0.6\textwidth]{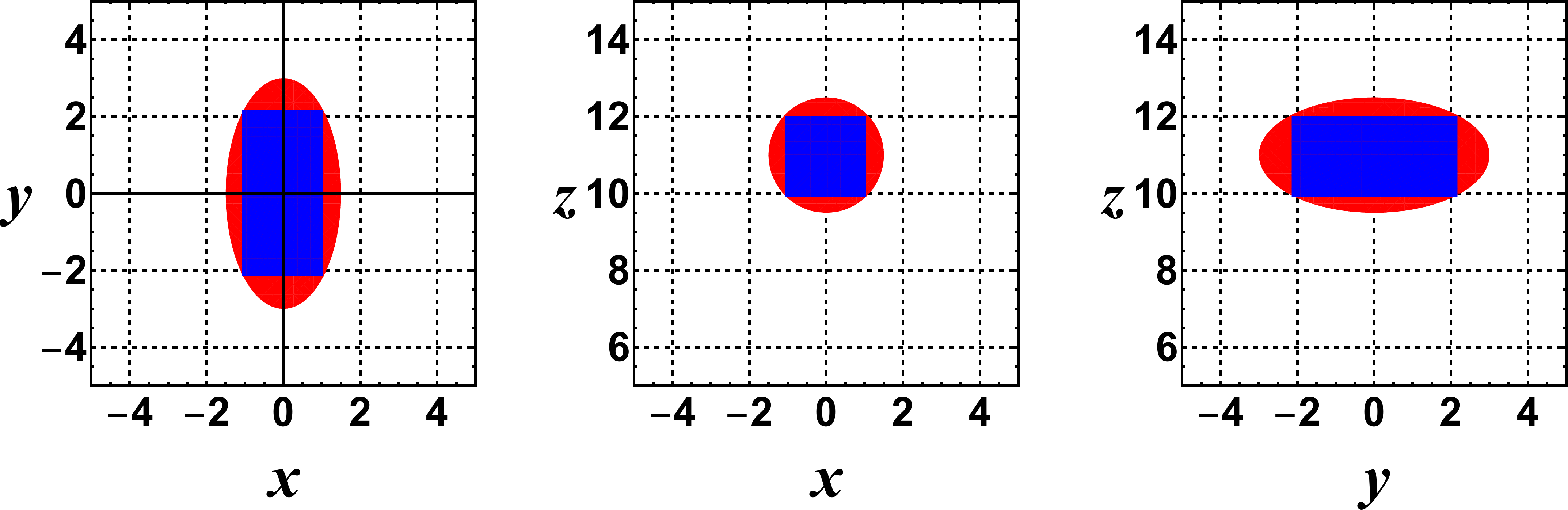}
\caption{
Same as Fig.~\ref{fig:ellipsoid1} but cross sections are shown. The cuboid (blue) and ellipsoid (red) are shown, from the left, on the plane at $z=11$, on the $x$-$z$ plane, and on the $y$-$z$ plane, respectively.
}
\label{fig:ellipsoid2}
\end{figure}

(Step 3) Now we give $\vv{a}_0^{\rm cuboid}$ from $\vv{a}_*^{\rm cubic}$ as
$\vv{a}_0^{\rm cuboid}=(x_1,x_2,y_1,y_2,z_1,z_2,M)=(x_0'-l'/2,x_0'+l'/2,y_0'-l'/2,y_0'+l'/2,z_0'-l'/2,z_0'+l'/2,M')$. Then assuming a cuboid, we obtain reconstructed values $\vv{a}_*^{\rm cuboid}$ as
\be
\vv{a}_*^{\rm cuboid}=(x_1^*,x_2^*,y_1^*,y_2^*,z_1^*,z_2^*,M^*)=
(-1.074,\,1.036,\,-2.146,\,2.166,\,9.908,\,12.02,\,0.029).
\ee
The reconstructed cuboid is shown in Fig.~\ref{fig:ellipsoid1} with the ellipsoidal target. The cuboid (blue) and ellipsoid (red) are shown in two-dimensional planes in Fig.~\ref{fig:ellipsoid2}. We emphasize that the narrowing process $\Gamma\to\vv{a}_0^{\rm cubic}\to\vv{a}_0^{\rm cuboid}$ is essential. If $\vv{a}_0^{\rm cubic}$ is used for the initial guess for the cuboid tomography, the iteration of the Levenberg-Marquardt method does not converge except for some special cases.

\section{Beef experiment}
\label{beefexp}

\begin{figure}[htbp]
\centering
\includegraphics[width=0.4\textwidth]{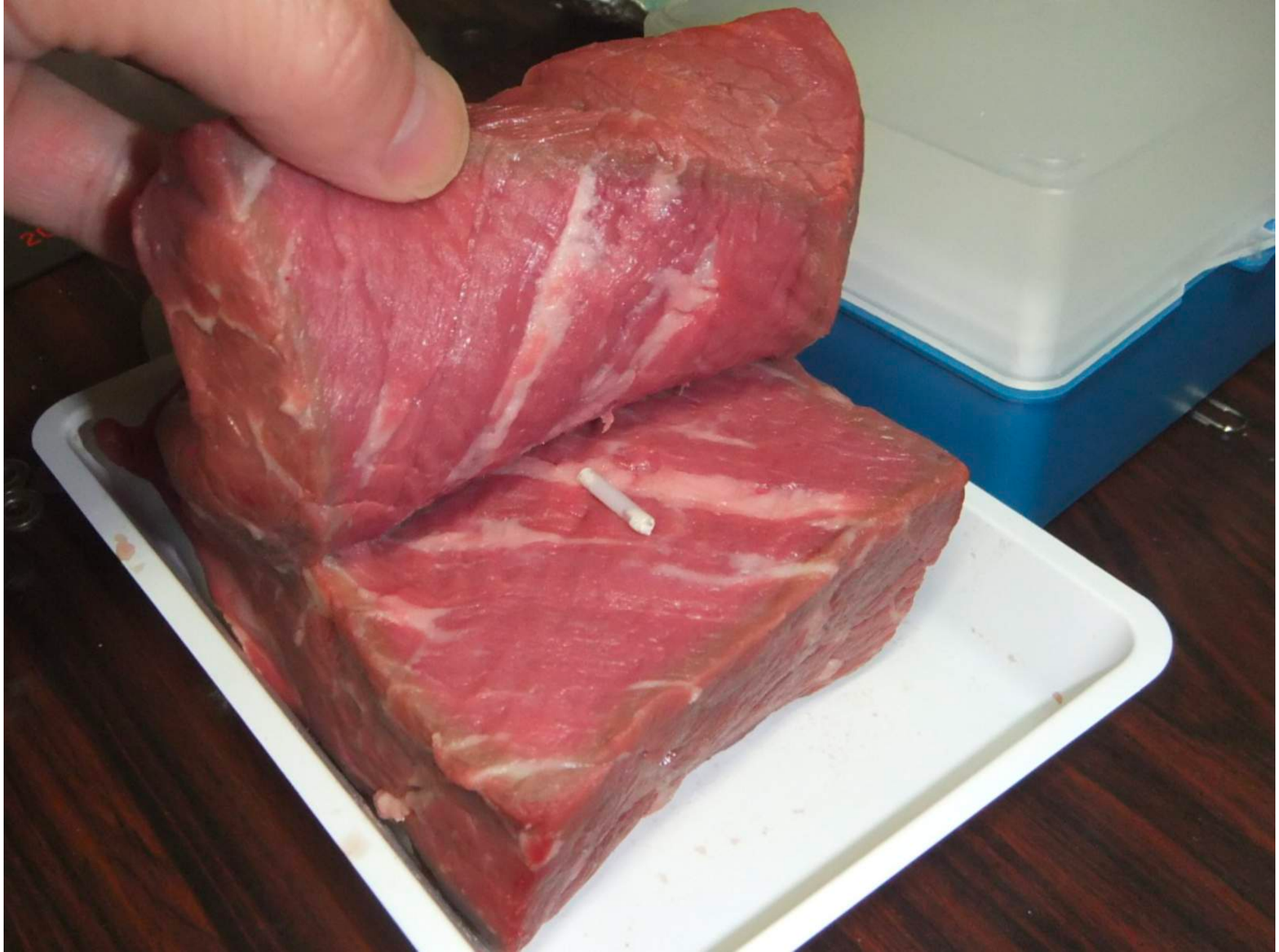}
\hspace{5mm}
\includegraphics[width=0.4\textwidth]{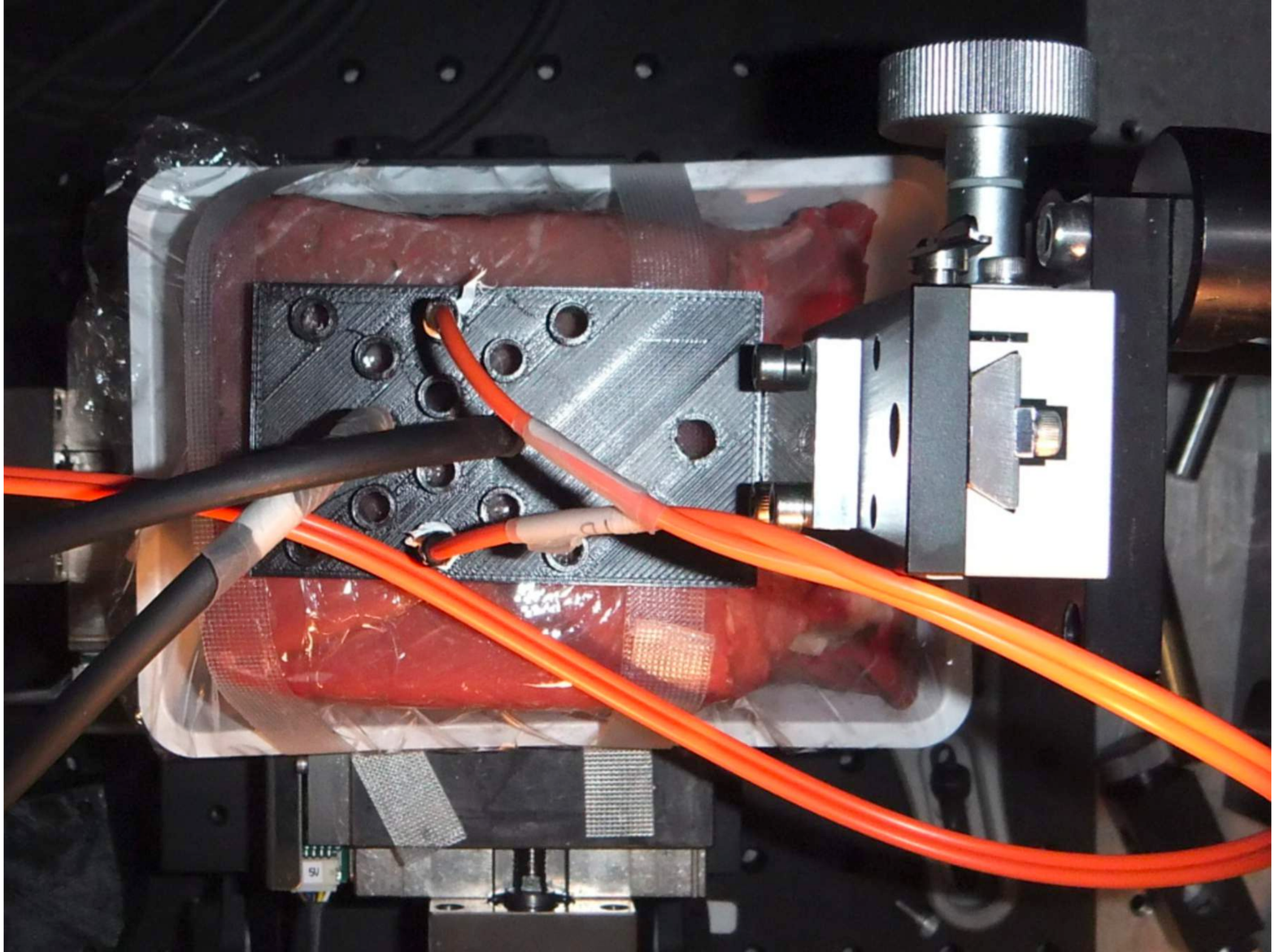}
\caption{
(Left) A tube of ICG was placed inside the beef. (Right) Boundary measurements were performed using optical fibers attached to a holder on the top of the beef.
}
\label{fig:beef}
\end{figure}

Let us reconstruct a tube which contains ICG in beef with our reconstruction scheme. Figure \ref{fig:beef} shows how the tube was placed in the beef. The tube has the shape of a cylinder of length $8\,{\rm mm}$ and diameter $2\,{\rm mm}$. We performed time-dependent measurements using a holder placed on the top of the beef. Four optical fibers (two are for sources and the other two are for detectors) are attached to the holder. The table on which the beef is placed changes positions while the holder is fixed.

Optical parameters for the beef are $\mu_s'=0.92\,{\rm mm}^{-1}$, $\mu_a =0.023\,{\rm mm}^{-1}$. The refractive index is set to $1.37$. Moreover, $\tau=0.6\,{\rm ns}$. After subtracting the background fluorescence, we could use $N_{\rm SD}=16$ source-detector pairs $p_i=(\vv{r}_s^i,\vv{r}_d^i)$ ($i=1,\dots,N_{\rm SD}$):
$p_{1}=(-5-10\sqrt{3},0;-5,10)$, $p_{2}=(-5+10\sqrt{3},0;-5,10)$,
$p_{3}=(-5-10\sqrt{3},5;-5,15)$, $p_{4}=(-5+10\sqrt{3},5;-5,15)$,
$p_{5}=(-10\sqrt{3},0;0,10)$, $p_{6}=(-10\sqrt{3},5;0,15)$,
$p_{7}=(5-10\sqrt{3},5;5,-5)$, $p_{8}=(5-10\sqrt{3},5;5,15)$,
$p_{9}=(5-10\sqrt{3},0;5,10)$, $p_{10}=(5-10\sqrt{3},-5;5,5)$,
$p_{11}=(-10\sqrt{3},-5;0,5)$, $p_{12}=(-10+10\sqrt{3},0;-10,10)$,
$p_{13}=(-15+10\sqrt{3},0;-15,10)$, $p_{14}=(-15+10\sqrt{3},5;-15,-5)$,
$p_{15}=(-15+10\sqrt{3},5;-15,15)$, $p_{16}=(-10+10\sqrt{3},5;-10,15)$, 
where the unit of coordinates is ${\rm mm}$. We set $N_t=20$ and $\Delta t=6.1\,{\rm ps}$. The peak of $U_m^{\rm exp}(\vv{r}_d^i,t;\vv{r}_s^i)$ comes at $t=(k_0^i-1)\Delta t$ ($i=1,\dots,N_{\rm SD}$). Measured times used for reconstruction are $t_k^i=t_0^i+(k-1)\Delta t$ ($k=1,\dots,N_t$), where $t_0^i=(k_0^i-11)\Delta t$.

First we set $\Gamma=\{-10<x<0,5<y<20\}$ since measured values were large in this region such as $I_5=3.3\times10^4$, $I_6=4.0\times10^4$, $I_{12}=2.9\times10^4$, $I_{16}=3.4\times10^4$. Although the value of $n(\vv{r})$, i.e., $M$ for the cube and cuboid, is a parameter to be reconstructed, $M$ is determined only up to a constant which comes from the property of the detector. We set $(x_0,y_0,z_0)=(-5,10,7)$, $l=2$, and obtain
\be
(x_0',y_0',z_0')=(-4.12,\,7.72,\,17,25),\quad l'=3.92.
\ee
With the above values as the initial guess we perform the iterative method once again. The reconstructed values are obtained as
\be
x_1^*=-5.16,\quad x_2^*=-3.11,\quad y_1^*=3.83,\quad y_2^*=12.03,\quad
z_1^*=16.05,\quad z_2^*=16.34.
\ee
The position of the cuboid is what we expected. For example, we see that the orientation of the cylinder should be almost parallel to the $y$-axis. Thus the position of the fluorescence tube is identified with the proposed numerical scheme.

\section{Conclusion}
\label{concl}

By giving up shape reconstruction, we can identify the location of the target by reconstructing only several unknown parameters. Even for these several parameters, the Levenberg-Marquardt algorithm is not stable unless good initial guesses are used. Thus the proposed procedure of narrowing target domains as $\Gamma\to\vv{a}_0^{\rm cubic}\to\vv{a}_0^{\rm cuboid}$ is important for the iterative method to work.

Although we assumed the half space in this paper and made use of an analytical solution to the diffusion equation, the proposed algorithm works also in more general cases where diffusion equations must be solved numerically by finite difference method or finite element method \cite{Zhu-etal11}. The proposed algorithm can be applied not only to the Levenberg-Marquardt method but also to other iterative schemes such as the conjugate gradient method and the Gauss-Newton method.

In this paper, the algorithm was explained using a single target. The generalization of the method for multiple targets is straightforward at least if the number of targets is known.

\section*{Funding}

NSFC (No.11421110002, No.11531005, No.91730304) (to J. J. Liu); Grant-in-Aid for Scientific Research (17K05572 and 17H02081) of the Japan Society for the Promotion of Science (JSPS) (to M. Machida); Grant-in-Aid for Scientific Research (15K21766 and 15H05740) of JSPS (to G. Nakamura). Also, the JSPS A3 foresight program: Modeling and Computation of Applied Inverse Problems.

\section*{Disclosures}

The authors declare that there are no conflicts of interest related to this article.



\begin{thebibliography}{1}
\newcommand{\enquote}[1]{``#1''}

\bibitem{Corlu-etal07}
A. Corlu, R. Choe, T. Durduran, M. A. Rosen, M. Schweiger, S. R. Arridge, M. D. Schnall, and A. G. Yodh,
\enquote{Three-dimensional in vivo fluorescence diffuse optical tomography of breast cancer in humans,}
{\protect\JournalTitle{Optics Express}} \textbf{15}, 6696--6716 (2007).

\bibitem{Ducros-DAndrea-Bassi-Peyrin11}
N. Ducros, C. D'Andrea, A. Bassi, and F. Peyrin,
\enquote{Fluorescence diffuse optical tomography: Time-resolved versus continuous-wave in the reflectance configuration,}
{\protect\JournalTitle{IRBM}} \textbf{32}, 243--250 (2011).

\bibitem{Gao-Zhao-Zhang-Tanikawa-Marjono-Yamada08}
F. Gao, H. Zhao, L. Zhang, Y. Tanikawa, A. Marjono, and Y. Yamada,
\enquote{A self-normalized, full time-resolved method for fluorescence diffuse optical tomography,}
{\protect\JournalTitle{Optics Express}} \textbf{16}, 13104--13121 (2008).

\bibitem{Koenig-etal08}
A. Koenig, L. Herv\'{e}, V. Josserand, M. Berger, J. Boutet, A. Da Silva, J.-M. Dinten, P. Pelti\'{e}, J.-L. Coll, P. Rizo,
\enquote{In vivo mice lung tumor follow-up with fluorescence diffuse optical tomography,}
{\protect\JournalTitle{J. Biomed. Opt.}} \textbf{13}, 011008 (2008).

\bibitem{Levenberg44}
K. Levenberg,
\enquote{A Method for the solution of certain non-linear problems in least squares,}
{\protect\JournalTitle{Quarterly Appl. Math.}} \textbf{2}, 164--168 (1944).

\bibitem{Marquardt63}
D. W. Marquardt,
\enquote{An algorithm for least-squares estimation of nonlinear parameters,}
{\protect\JournalTitle{J. Soc. Indust. Appl. Math.}} \textbf{11}, 431--441 (1963).

\bibitem{Ntziachristos-etal02}
V. Ntziachristos, C. H. Tung, C. Bremer, and R. Weissleder,
\enquote{Fluorescence molecular tomography resolves protease activity in vivo,}
{\protect\JournalTitle{Nature Medicine}} \textbf{8}, 757--761 (2002).

\bibitem{Ntziachristos-Bremer-Weissleder03}
V. Ntziachristos, C. Bremer, and R. Weissleder,
\enquote{Fluorescence imaging with near-infrared light: new technological advances that enable in vivo molecular imaging,}
{\protect\JournalTitle{Eur. Radiol.}} \textbf{13}, 195--208 (2003).

\bibitem{Ntziachristos-Ripoll-Wang-Weissleder05}
V. Ntziachristos, J. Ripoll, L. V. Wang, and R. Weissleder,
\enquote{Looking and listening to light: the evolution of whole-body photonic imaging,}
{\protect\JournalTitle{Nature Biotech.}} \textbf{23}, 313--320 (2005).

\bibitem{Panasyuk-Wang-Schotland-Markel08}
G. Y. Panasyuk, Z.-M. Wang, J. C. Schotland, and V. A. Markel,
\enquote{Fluorescent optical tomography with large data sets,}
{\protect\JournalTitle{Opt. Lett.}} \textbf{33}, 1744--1746 (2008).

\bibitem{Zhu-etal11}
Q. Zhu, H. Dehghani, K. M. Tichauer, R. W. Holt, K. Vishwanath, F. Leblond, and B. W. Pogue,
\enquote{A three-dimensional finite element model and image reconstruction algorithm for time-domain fluorescence imaging in highly scattering media,}
{\protect\JournalTitle{Phys. Med. Biol.}} \textbf{56}, 7419--7434 (2011).

\end{thebibliography}
\end{document}